\documentclass[traditabstract]{aa}

\usepackage{natbib}
\usepackage{graphicx}
\bibpunct{(}{)}{;}{a}{}{,}
\usepackage{txfonts}

\begin{document}

\title{Detection of Six Rapidly Scintillating AGNs and the Diminished Variability of J1819+3845}

\author{J. Y. Koay \inst{1}
\and H. E. Bignall \inst{1}
\and J.-P. Macquart \inst{1}
\and D. L. Jauncey \inst{2,3}
\and B. J. Rickett \inst{4}
\and J. E. J. Lovell \inst{5}}

\institute{International Centre for Radio Astronomy Research, Curtin University, Bentley, WA 6102, Australia
\and CSIRO Astronomy and Space Science, Australia Telescope National Facility, Epping, NSW 1710, Australia
\and Mount Stromlo Observatory, Weston, ACT 2611, Australia
\and Department of Electrical and Computer Engineering, University of California, San Diego, La Jolla, CA 92093, USA
\and School of Mathematics and Physics, University of Tasmania, TAS 7001, Australia}

\date{Received ... / Accepted ...}

\abstract {The extreme, intra-hour and $> 10\%$ rms flux density scintillation observed in AGNs such as PKS 0405-385, J1819+3845 and PKS 1257-326 at cm wavelengths has been attributed to scattering in highly turbulent, nearby regions in the interstellar medium. Such behavior has been found to be rare. We searched for rapid scintillators among 128 flat spectrum AGNs and analyzed their properties to determine the origin of such rapid and large amplitude radio scintillation. The sources were observed at the VLA at 4.9 and 8.4 GHz simultaneously at two hour intervals over 11 days. We detected six rapid scintillators with characteristic time-scales of $<$ 2 hours, none of which have rms variations $> 10\%$. We found strong lines of evidence linking rapid scintillation to the presence of nearby scattering regions, estimated to be $< 12$ pc away for $\sim 200$ $\mu$as sources and $<$ 250 pc away for $\sim 10$ $\mu$as sources. We attribute the scarcity of rapid \textit{and} large amplitude scintillators to the requirement of additional constraints, including large source compact fractions. J1819+3845 was found to display $\sim 2\%$ rms variations at $\sim 6$ hour time-scales superposed on longer $>$ 11 day variations, suggesting that the highly turbulent cloud responsible for its extreme scintillation has moved away, with its scintillation now caused by a more distant screen $\approx$ 50 to 150 pc away.}


\titlerunning{Six Rapid Scintillators and Diminished Scintillation in J1819+3845}

\maketitle
   
   \section{Introduction}\label{introduction}
   
Hourly variations in the flux densities of the most compact active galactic nuclei (AGNs) at cm wavelengths have incontrovertibly been shown to result from interstellar scintillation (ISS). The evidence comes from observations of time delays between variability patterns across widely-separated telescopes \citep{jaunceyetal00, dennett-thorpedebruyn02, bignalletal06}, as well as annual cycles in their variability timescales \citep{rickettetal01, jaunceymacquart01, bignalletal03, dennett-thorpedebruyn03, jaunceyetal03} consistent with relative motion between the scattering region and the Earth. These observations were made possible through the monitoring of the so called `extreme scintillators' that display rms flux density variations $>$ 10\% on timescales $\lesssim$ 1 hour, of which PKS 0405-385 \citep{kedziora-chudczeretal97}, J1819+3845 \citep{dennett-thorpedebruyn00} and PKS 1257-326 \citep{bignalletal03} are the most well-known. Such rapid and large amplitude ISS has been attributed to the presence of nearby turbulent clouds, $< 30\,{\rm pc}$ from the Sun.

These extreme scintillators are surprisingly rare, considering that the three archetypal sources were detected serendipitously. In the 5 GHz, Microarcsecond Scintillation Induced Variability (MASIV) Survey of 443 compact flat-spectrum sources \citep{lovelletal08}, no new examples (apart from J1819+3845) were found to display sustained modulation indices (rms variations as a percentage of mean flux density) of more than 10\%, although 16\% of the sources were found to scintillate on timescales of $< 12$ hours. This implied that the nearby, turbulent clouds cover only a small fraction of the sky. It also begs the question as to whether the presence of these nearby clouds are the only necessary condition for rapid, large amplitude ISS, or if there are other factors that contribute to their scarcity (ie. source compactness). 

J1819+3845 was the most extreme of this class. During a 4.8 GHz, 6-hour monitoring program on the European VLBI Network (EVN) in 2008 March, its extreme scintillation sustained over at least eight years was found to have ceased \citep{cimo08}, with an upper limit of 1\% rms variations (Cim\`{o}, private communication). Understanding the unexpected disappearance of the extreme variability of this source will shed some light on its origin at earlier epochs. We note that episodic extreme scintillation has been observed in PKS 0405-385 \citep{kedziora-chudczer06}, attributed either to the repeated appearance and expansion of new components in the source or spatially intermittent turbulent patches in the ISM drifting across its sight-line. 

In this letter, we examine data from a dual-frequency survey of ISS in 128 flat-spectrum AGNs to search for rapid scintillators, then make use of their properties to understand the origin of rapid radio scintillation. We present a method for identifying rapid scintillators and include a list of candidates in Section~\ref{rapid}. In Section~\ref{j1819}, we report on the variability of J1819+3845 subsequent to the period of extreme scintillation. We then discuss the implications of our results on the physics behind the most extreme scintillators in Section~\ref{discussion}.
        
	\section{Detection of the Six Rapid Scintillators}\label{rapid}
	
	 \begin{figure}
     \resizebox{\hsize}{!}{\includegraphics{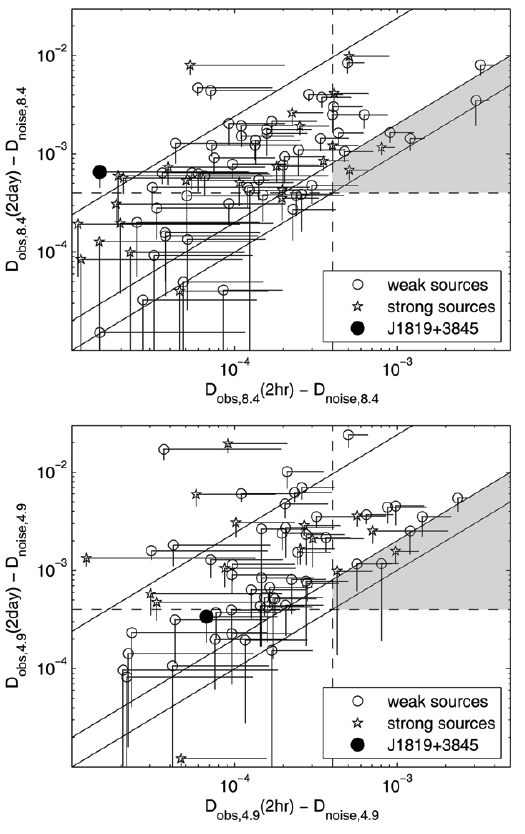}}
     \caption{$D_{obs}({\rm 2\,day}) - D_{noise}$ plotted against $D_{obs}{\rm (2\,hr)} - D_{noise}$ at 8.4 GHz (top) and 4.9 GHz (bottom). For each source, only the lower limits of the 1$\sigma$ errors in $D({\rm 2\,day})$ and the upper limits of the 1$\sigma$ errors in $D({\rm 2\,hr})$ are shown. The solid lines have a slope of 24, 2 and 1 (from top to bottom) respectively. The dashed lines represent the $5\sigma$ thresholds. Rapid scintillators are located within the shaded regions.} 
     \label{D2dayvsd2hr}
     \end{figure}
     	
As a follow-up to the MASIV Survey, a sub-sample containing 128 of the 443 original MASIV sources was observed at the VLA in 2009 January \citep[][ hereafter Paper I]{koayetal11}. The increased sensitivity from the larger number of antennas per subarray and observation span of 11 days enabled the measurement of rms flux density variations down to a level of $\sim$1\% and the estimation of scintillation timescales. Simultaneous observations at 4.9 GHz and 8.4 GHz on two subarrays provided a robust means of discriminating flux variations of astrophysical origin from variations due to instrumental and systematic errors.
          
	We characterized the variability of each source using the discrete structure function (SF) of the lightcurves, given by:
\begin{equation}\label{structurefunction}
D_{obs}(\tau)=\frac{1}{N_{\tau}}\sum_{j,k}[S(t_{j})-S(t_{k}-\tau)]^{2},
\end{equation}
where $S(t)$ is the flux density of the source at time $t$, normalized by its mean flux density. $N_{\tau}$ is the number of pairs of flux densities with a time-lag $\tau$. We also estimated a constant $D_{noise}$ for each source, which accounts for variability arising from instrumental and systematic errors including noise, low-level confusion, as well as residual gain and calibration errors, which we then subtract from $D_{obs}(\tau)$ to obtain the `true variability' of each source, $D(\tau)$. The details are described in Paper I.
	
	We found significant variability in 30\% of the sources at 2-hour time-scales, the shortest time interval between successive observations of each source. These sources have $D({\rm 2\,hr}) > 2 \times 10^{-4}$ on at least one frequency, so that $D_{obs}({\rm 2\,hr}) > 3\sigma$ above $D_{noise}$ ($\approx 1 \times 10^{-4}$ on average at both frequencies). These sources also display strongly correlated variability at both frequencies, based on the cross-correlation function, and a visual inspection of the lightcurves. Such correlations are unlikely to result from antenna-based or array-based errors as they were observed on two separate subarrays. Neither are these variations likely to be confusion, since the subarrays each have a different uv-coverage arising from dissimilar primary beams and synthesized beams at the two frequencies. On the other hand, flux variations due to weak ISS are expected to be correlated across a wide-bandwidth \citep{narayan92}. We also found larger fractions of sources displaying significant 2-hour variability at 4.9 GHz and in the weak sample of sources ($S_{\nu} < 0.3$ Jy), consistent with a population of brightness temperature limited sources whose variability is dominated by weak ISS \citep{lovelletal08}. No similar variations were observed in sources close to each other, ruling out atmospheric effects. None of the sources were observed at low solar elongations, so interplanetary scintillation is also negligible.   
	
	We devised a method to distinguish the rapid scintillators with characteristic timescales of $\tau_{char} < 2$ hours from the sources with much longer $\tau_{char}$ that display significant 2-hour variability. We have defined $\tau_{char}$ as the time-scale at which $D(\tau)$ reaches half its value at saturation. $D(\tau)$ increases with $\tau^{a}$ and saturates at twice the true variance of the source, when the source is observed for a duration much longer than $\tau_{char}$. Letting $\tau_{sat}$ be the time-scale at which $D(\tau)$ saturates, we obtain: 
\begin{equation}\label{ratio}
\frac{D({\rm 2\,day})}{D({\rm 2\,hr})}  \left\lbrace  
\begin{array}{l l}
   = 1 & \quad \mbox{if $\tau_{sat} \leq$ 2 hr}\\
   = (\tau_{sat}/\mbox{2 hr})^{a} & \quad \mbox{if 2 hr $< \tau_{sat} <$ 2 days}\\ 
   \geq 24^{a} & \quad \mbox{if $\tau_{sat} \geq$ 2 days}\\ 
  \end{array} 
  \right.
\end{equation}	
Figure~\ref{D2dayvsd2hr} plots $D_{obs}({\rm 2\,day}) - D_{noise}$ against $D_{obs}({\rm 2\,hr}) - D_{noise}$ at both frequencies so that $D({\rm 2\,day})/D({\rm 2\,hr})$ is represented by a slope. We determined that a source must satisfy both of the following conditions at at least one frequency to be classified as a rapid scintillator, corresponding to shaded regions in Figure~\ref{D2dayvsd2hr}:

1. $D({\rm 2\,day})/D({\rm 2\,hr}) < 2$, since $\tau_{char} \geq 0.5(\tau_{sat})$ for $a \geq 1$. We include also sources that lie outside this region, but have 1$\sigma$ error bars that extend into this region.

2. $D({\rm 2\,day})$ and $D({\rm 2\,hr})$ must be $> 4 \times 10^{-4}$, so that $D_{obs}(\tau) > 5\sigma$ above $D_{noise}$.
	
\begin{figure}
     \resizebox{\hsize}{!}{\includegraphics{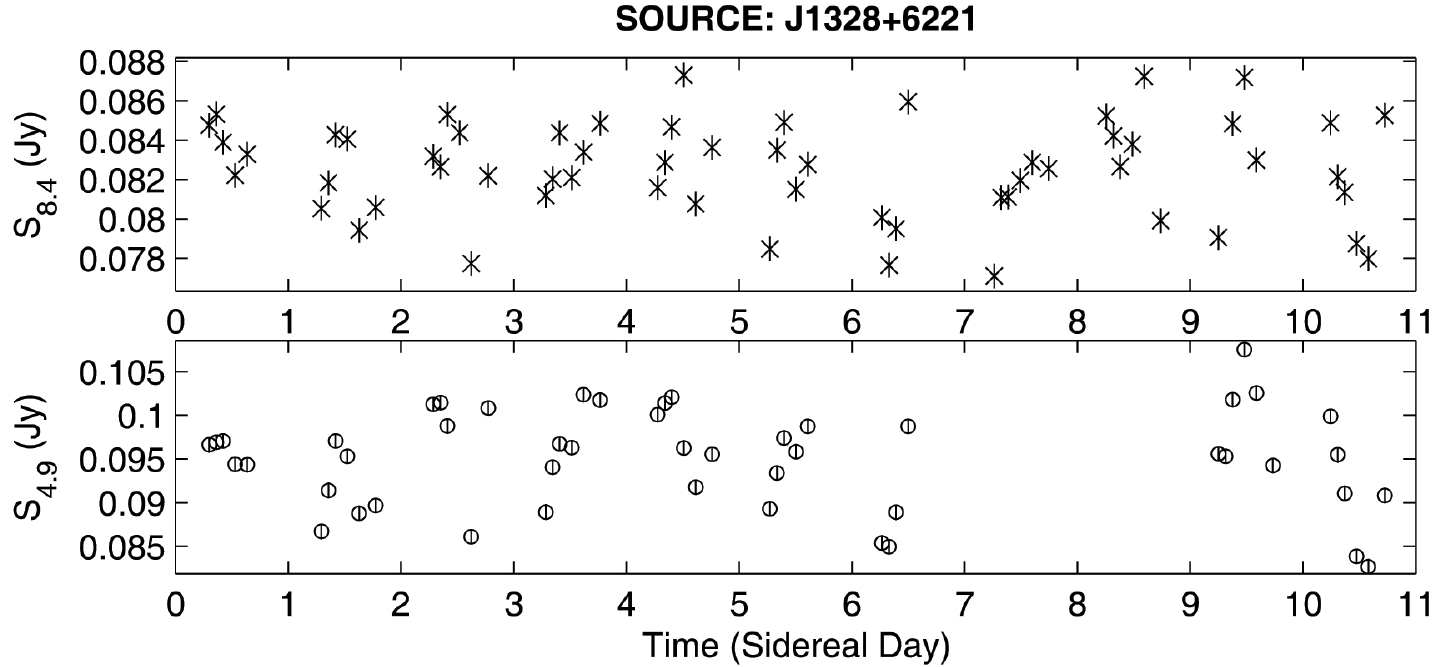}}
     \caption{Lightcurves for the rapid scintillator J1328+6221 at 8.4 GHz (top) and 4.9 GHz (bottom).}
     \label{1328}
     \end{figure}

	We detected six rapid scintillators in the sample (listed in Table~\ref{rapidsources}), out of an initial ten candidates that fulfilled the conditions above. Four candidates were found to be slower scintillators after a visual inspection; they contain outliers in the lightcurves that skew $D({\rm 2\,hr})$ towards larger values, or display quasi-periodic structure in the lightcurves leading to quasi-periodicity in the SFs, thereby reducing $D({\rm 2\,day})$ when 2 days is close to a multiple of the quasi-period. For the six confirmed rapid scintillators, a cross-check with their $\tau_{char}$ obtained using an independent method in Paper I revealed that they were also estimated to be $\lesssim 2$ hours on at least one frequency. The lightcurves of these six sources show strong correlation across both frequencies. An example, J1328+6221, is shown in Figure~\ref{1328}.
	          
     \begin{table}
\centering
\caption{{Rapid Scintillators Detected in our Sample and Their Properties.} \label{rapidsources}}
\scriptsize
\begin{tabular}{c c c c c c c c}
\hline
\hline
Source Name & \multicolumn{2}{c}{Galactic Coordinates} & $I_{\alpha}$\tablefootmark{a} & $S_{4.9}$\tablefootmark{b} & $S_{8.4}$\tablefootmark{b}  & $m_{4.9}$\tablefootmark{c} & $m_{8.4}$\tablefootmark{c}\\
(J2000)	& $l$ & $b$ & (R) & (Jy) & (Jy) & (\%) & (\%)\\
\hline

J0800+4854	& 170.11 &  31.16 & 0.5 & 0.10 & 0.08 & 5.0 & 5.1\\	
J0929+5013	& 167.66 &  45.42 & 0.6 & 0.40 & 0.39 & 4.1 & 2.6\\	
J1049+1429  & 230.79 &  59.00 & 0.9 & 0.13 & 0.15 & 3.0 & 2.8\\		
J1328+6221	& 115.56 &  54.23 & 0.5 & 0.10 & 0.08 & 5.9 & 3.0\\	
J1549+5038  & 80.24  &  49.06 & 0.4 & 0.91 & 0.93 & 3.0 & 2.1\\	
J1931+4743	& 79.98  &  13.53 & 5.2 & 0.11 & 0.10 & 7.2 & 7.4\\	
\hline
\end{tabular}
\tablefoot{\tablefoottext{a}{Line-of-sight H$\alpha$ intensity in units of Rayleighs (R), from \citet{haffneretal03}}
\tablefoottext{b}{2009 January VLA mean flux density}
\tablefoottext{c}{2009 January raw modulation index, without error subtraction}}
\end{table} 
           
   \section{The Variability of J1819+3845}\label{j1819}
   
\begin{figure}
     \resizebox{\hsize}{!}{\includegraphics{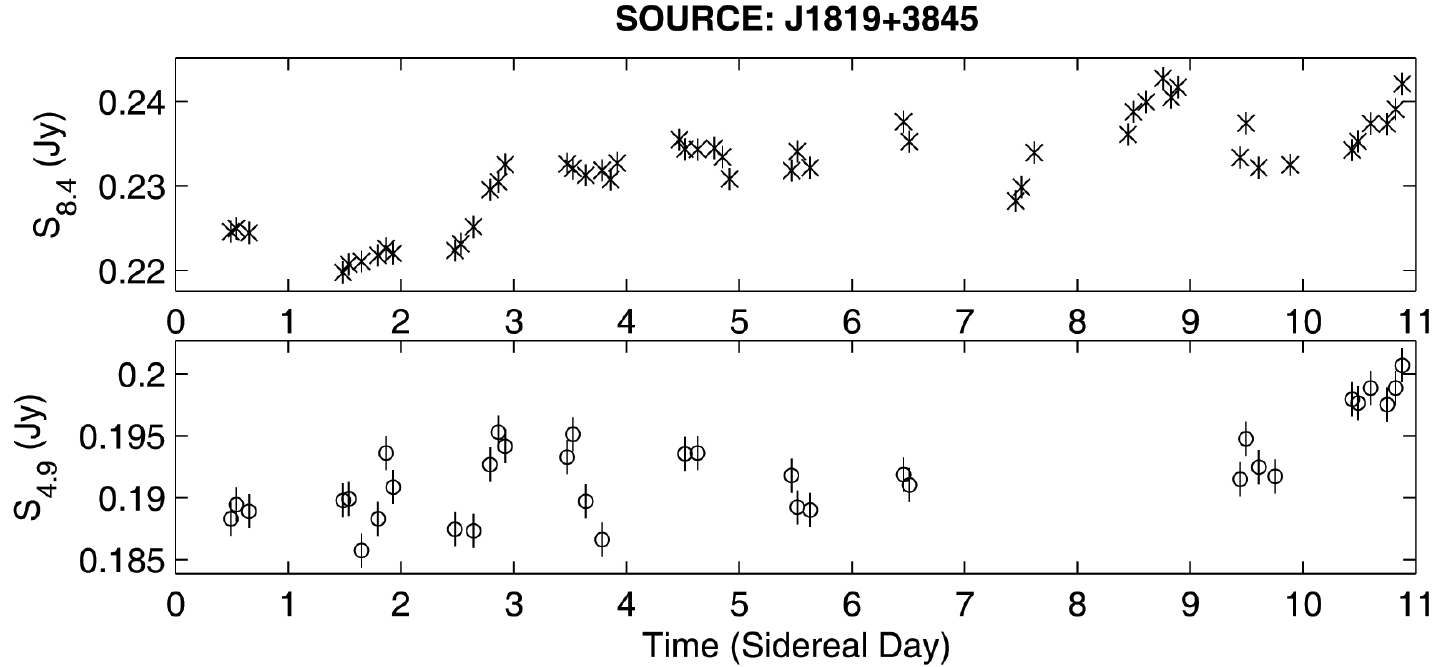}}
     \caption{Lightcurves for the former extreme scintillator J1819+3845 at 8.4 GHz (top) and 4.9 GHz (bottom).} 
     \label{1819}
     \end{figure}

   Our observations confirm that the character of the rapid and large amplitude scintillation in J1819+3845 has changed significantly (it is no longer exhibiting extreme scintillation, see Figures~\ref{D2dayvsd2hr} and \ref{1819}), as previously noted by \citet{cimo08}. We detected rms variations of only 1.5\% at 4.9 GHz and 2.5\% at 8.4 GHz. Variations of $\tau_{char} \sim 6$ hours dominate at 4.9 GHz, while the variability at 8.4 GHz is dominated by the slow rise in flux density over the 11-day period, also discernible in the 4.9 GHz lightcurve. Such variations would not have been detected in the 6-hour observations by Cim\`{o}. 
    
   There are two possible explanations for the factor of $\sim 20$ decrease in scintillation amplitude and the factor of $\sim 12$ increase in timescale from the previously observed rms variations of 25-40\% and time-scale of 30 minutes \citep{macquartdebruyn07}. Either the source has expanded, thereby quenching the scintillation, or the highly turbulent patch in the Local Interstellar Medium (LISM) responsible for the extreme scintillation has drifted off its sight-line. If the former is true, it would require the apparent angular size of the source to have expanded by a factor of 12 in $<$ 2 years from 2006 to 2008. For a $>100\,\mu$as source (65 pc at its measured redshift of 0.54, $H_{0} = 70{\,\rm kms^{-1} Mpc^{-1}}$, $\Omega_{M} = 0.27$ and $\Omega_{\Lambda} = 0.73$), this requires an apparent expansion speed of $> 5.8c$. As there is no discernible change in its mean flux density, and the source remains unresolved at mas scales at all EVN baselines which includes the Urumqi telescope \citep{cimo08}, such an explanation is unlikely. 
   
   The latter explanation is therefore more likely. Assuming that the source size has not changed and that the velocity of the screen is $50\, {\rm kms}^{-1}$, we estimate that the scattering region now lies at a distance of 50 to 150 pc away, requiring also a factor $\sim 9$ decrease in the length-scale, $r_{\rm diff}$, of the phase fluctuations at the scattering screen to achieve the observed reduction in the modulation index (see Equation~\ref{modindex}). The slower, $\tau_{sat} >$ 11 day variations could be intrinsic to the source, or ISS at a second, more distant screen $\sim 1.7$ kpc away. In any case, these longer time-scale variations would have been masked by the previous rapid and large amplitude variations of J1819+3845.

   \section{The Origin of Rapid and Extreme Scintillation}\label{discussion}

   The spatial distribution of the six rapid scintillators shows a dependence on Galactic latitude and line-of-sight free electron content, consistent with the view that the most rapid ISS in quasars is caused by scattering in nearby regions. Of the six sources, five of them lie at Galactic latitudes of $> 30^{\circ}$, and have H$\alpha$ intensities obtained from the Wisconsin H-Alpha Mapper Northern Sky Survey \citep{haffneretal03} of $I_{\alpha}\leq 1.0$ Rayleighs, as a proxy for the line-of-sight integral of the square of the free electron density. In fact, out of 20 sources in our sample that have $I_{\alpha} \leq 0.6$ Rayleighs, of which 14 have variability amplitudes 3$\sigma$ above $D_{noise}$, four of them are rapid scintillators. On the other hand, in 108 sources with $I_{\alpha} > 0.6$ Rayleighs, we find only two rapid scintillators. Fisher's Exact Test for contingency tables confirms that the association between rapid ISS and low $I_{\alpha}$ is statistically significant at the 0.05 level. We calculated a probability of 0.0044 that the correlation was obtained purely by chance when considering the entire sample of 128 sources, and a probability of 0.0026 when considering only the $> 3\sigma$ variable sources. The significance holds even if we consider J1819+3845 ($I_{\alpha} = 2.2$ Rayleighs) as a rapid scintillator based on its history, and include PKS 0405-385 and PKS 1257-326 in our sample, which have $I_{\alpha} < 0.5$ and $I_{\alpha} = 16$ Rayleighs respectively (obtained from the Southern H-Alpha Sky Survey Atlas, see \citet{gaustadetal01}). This is consistent with the MASIV results which show that the fraction of fast scintillators ($\tau_{char} < 0.5$ days) decreases with increasing H$\alpha$ intensity \citep{lovelletal08}. These sight-lines with low H$\alpha$ at higher Galactic latitudes correspond to regions of lower transition frequencies ($\nu_{0}$) between weak and strong ISS \citep{walker98}, which imply lower effective scattering screen distances at a fixed observing frequency. These nearby screens produce smaller scale interference patterns on the Earth's surface, leading to more rapid scintillation for a given screen velocity as they drift across the telescope.
   
   The $\tau_{char} <$ 2 hours can be attributed to scattering at an effective screen distance of $L <$ 12 pc for a $\sim 200 \, \mu$as source component, or at $L <$ 250 pc for a $\sim 10 \, \mu$as component, for typical transverse velocities of $50 {\,\rm kms}^{-1}$. Assuming Kolmogorov turbulence and $\nu_{0} = 3$ GHz, numerical computations using the fitting formula for ISS in \citet{goodmannarayan06} indicate that the former will produce the observed $\sim 5\%$ rms variations, giving brightness temperatures, $T_{B}$, of $4 \times 10^{10}$ K if all the 100 mJy flux density is contained within the $\sim 200 \,\mu$as component. The latter requires that the ratio of the $\sim 10 \,\mu$as component flux density to the total observed flux density (ie. the source compact fraction, $f_{c}$) be $\sim 25\%$ to obtain $\sim 5\%$ rms variations, giving $T_{B} = 4 \times 10^{12}$ K for an observed 100 mJy source. This is because the observed modulation index is normalized by the flux density measured at the resolution of the telescope, which may include emission from extended, non-scintillating components. 
   
   There are strong indications that both scenarios for rapid ISS occur in our sample. Gaussian model fits to source core components in 8.4 GHz VLBI images \citep{ojhaetal04} estimate core diameters of $\sim 200 \,\mu$as ($L <$ 12 pc) in J1328+6221 and J1931+4743, as well as $\sim 40 \,\mu$as ($L <$ 60 pc) in J1049+1429. The VLBI cores may contain an even more compact, unresolved component which could push the maximum screen distances further out. However, \citet{linskyetal08} have also shown that the extreme scintillation in J1819+3845, PKS 0405-385 and PKS 1257-326 may be associated with regions where multiple warm-ionized clouds $<$ 15 pc from the Sun intersect, and possibly interact to form highly-turbulent screens. We found that J1328+6221 and J1931+4743 also have sight-lines through (or at the edge of) these multiple clouds, providing further evidence for screens $<$ 15 pc. On the other hand, the sight-lines of J1049+1429 and the other three sources do not intersect these regions, suggesting larger screen distances.
   
   That rapid ISS arises from a large range of source sizes and scattering screen distances explains why rapid scintillators are not as rare as initially thought; it is the sources that display \textit{both} rapid \textit{and} large rms variations $>$ 10\% that are rare. The six rapid scintillators constitute 5\% of our sources, yet none have rms variations $>$ 10\%. No new extreme scintillators were found in the MASIV Survey \citep{lovelletal08}, although J0949+5819 and J0829+4018 did show $\sim 15\%$ rms variations at one epoch.  
   
   We argue that the scarcity of the extreme scintillators can be attributed to additional constraints required for rapid scintillators to also display rms variations $>$ 10\%. In the regime of weak ISS where the observing frequency, $\nu > \nu_{0}$, the modulation index for an extended source whose angular size, $\theta_{s}$, is larger than the Fresnel angular-scale at the scattering screen, $\theta_{F}$, is given by \citet{narayan92} and \citet{walker98}, which we rewrite as follows:
\begin{equation}\label{modindex}
m = f_{c}{\left( {\frac{\nu_{0}}{\nu}}\right)}^{\frac{17}{12}} {\left( {\frac{\theta_{F}}{\theta_{s}}} \right) }^{\frac{7}{6}} = \frac{c f_{c}}{2 \pi \nu} {\left( \frac{1}{r_{\rm diff}} \right)}^{\frac{5}{6}} {\left( \frac{1}{L} \right)}^{\frac{1}{6}} {\left( \frac{1}{\theta_{s}} \right)}^{\frac{7}{6}},
\end{equation}  
where $\theta_{F} \approx r_{\rm F}/L = \sqrt{c/(2\pi \nu L)}$, $r_{\rm F}$ is the Fresnel length-scale, $r_{\rm diff}$ is the diffractive length-scale determined by the strength of the turbulence at the scattering screen, and $r_{\rm F}/r_{\rm diff} = (\nu_0/\nu)^{17/10}$. Kolmogorov turbulence is assumed.

Firstly, the source must be observed at a frequency close to $\nu_{0}$, where $m$ for a point source $\sim$100\%. This is also true in the regime of strong refractive ISS ($\nu < \nu_{0}$), as $m$ decreases again with decreasing $\nu$. 

Secondly, the scattering screen must be highly turbulent (with small $r_{\rm diff}$). As discussed in Section~\ref{j1819}, such a turbulent cloud has possibly moved away in the case of J1819+3845. These turbulent patches may be localized and intermittent, possibly causing the episodic extreme scintillation in PKS0405-385 \citep{kedziora-chudczer06}. Such AU-scale inhomogeneities deviate from standard models of the ISM with homogenous Kolmogorov turbulence. They are instead reminiscent of the clumpy, discrete clouds observed in extreme scattering events \citep{fiedleretal87} and the scintillation of pulsar B0834+06 \citep{briskenetal10}. Scattering in such clouds would dominate the scintillation in PKS 1257-326 over any background ISS, despite its large line-of-sight $I_{\alpha}$. 

Thirdly, the source must have $f_{c}$ close to unity. Comparing VLBI flux densities of the core components to that of the extended mas structures \citep{ojhaetal04} gives upper limits to $f_{c}$ at VLBI scales; $\lesssim$ 75\% for J0800+4854, J1328+6221 and J1931+4743, $\lesssim$ 100\% for J1049+1429. J1549+5038 also shows significant extended structure at mas scales \citep{feycharlot00}. At the resolution of the VLA, $f_{c}$ will possibly be lower. This reduces the modulation indices in our six rapid scintillators. As an example, \citet{savolainenkovalev08} reported $m \sim$ 13\% scintillation at a time-scale of 2.7 hours in J1159+2914 with a mean flux density $\sim$ 1.5 Jy at 15 GHz, based on VLBI observations in 2007. The MASIV Survey and our observations found $m \sim$ 5\% at both frequencies, obtaining flux densities $\sim$ 3.0 Jy. The unnormalized peak to trough variations in the three separate experiments are in fact comparable in amplitude, $\approx$ 0.5 Jy.  
   
In summary, while scattering screens at distances up to 250 pc can produce $\tau_{char} \lesssim$ 2 hour ISS in AGN cores $\gtrsim$ 10 $\mu$as, source compact fractions $\gtrsim$ 50\% are necessary for $\sim$ 10 $\mu$as scintillating components ($\gtrsim$ 95\% for $\sim$ 100 $\mu$as components) to display $\gtrsim$ 20\% rms variations at $\tau_{char} \sim$ 1 hour when observed at $\nu = \nu_{0} \sim 4.9 \,\rm GHz$. We therefore predict that higher angular resolution observations will reveal more large amplitude rapid scintillators, providing more reliable constraints on the flux density of the scintillating components. 
   
   \begin{acknowledgements}
We thank the rest of the MASIV mob - Lucyna Kedziora-Chudczer, Tapio Pursimo, Roopesh Ojha and Cormac Reynolds for their involvement in the observations and helpful comments. We are also grateful to Giuseppe Cim\`{o} for providing information on his observations of J1819. JYK is supported by the Curtin Strategic International Research Scholarship. BJR thanks the National Science Foundation (NSF) for partial support under grant AST-0507713. We made use of data from the Wisconsin H-Alpha Mapper Survey and the Southern H-Alpha Sky Survey Atlas, both supported by the NSF, as well as data from the United States Naval Observatory Radio Reference Frame Image Database. The National Radio Astronomy Observatory is a facility of the NSF operated under cooperative agreement by Associated Universities, Inc. 
    \end{acknowledgements} 


\end{document}